\input psfig.sty
\rm
\magnification=1200
\baselineskip=12.5pt
\parskip=10pt plus 1pt
\parindent=0pt
\def\x{{\hat x}}
\def\y{{\hat y}}
\def\z{{\hat z}}
\null

\vskip .3truein
\centerline{\bf TEST EVOLUTION OF NON-AXISYMMETRIC}
\vskip .2truein
\centerline{\bf GRAVITATIONAL WAVES}

\vskip .4 truein
\leftskip 72pt{
\rm{Leonardo Sigalotti [1]  and Richard F. Stark [2]
\footnote{*}{E-mail address: richard@fiz.uni-lj.si 
                   {\sl and} norver@stark1.freeserve.co.uk} }\hfill\break
\sl{[1] International Center for Theoretical Physics,}\hfill\break
\sl{Strada Costiera 11, 34100 Trieste, Italy}\hfill\break
\sl{[2] University of Ljubljana, Department of Physics,}\hfill\break
\sl{Jadranska 19, 61000 Ljubljana, Slovenia}\hfill\break}

\leftskip 30pt
\vskip .4truein
{\bf Abstract:}\quad 
   We give a preliminary report on one of the tests we have performed
of a full non-axisymmetric general relativistic code.
The test considered here concerns the numerical
evolution of vacuum non-axisymmetric gravitational waves and their
comparison at low amplitudes with theoretical waveforms obtained
from linearised theory.


\leftskip 0pt
\parindent=30pt
\vskip .5 truein
\centerline {\bf I. \ Introduction}

   We are currently carrying out a number of detailed tests of
a full non-axisymmetric general relativistic code which solves 
the coupled Einstein and hydrodynamic equations using 
the `3+1' ADM formalism in the radial gauge with mixed slicing [1].  
The code has been written in particular to obtain directly the 
gravitational wave emission from non-axisymmetric rotating 
gravitational collapse.  
The test we report here concerns the numerical evolution of vacuum 
non-axisymmetric gravitational waves and their
comparison at low amplitudes with theoretical waveforms obtained 
using linearised theory.  This report should be considered
preliminary, since the code is still under active development.

   We consider the propagation of non-axisymmetric $l=2$ and $l=4$, 
and different $m$,
linearised gravitational waves.  We are able to use essentially
analytic predictions for these waves with which to compare to
the numerical solutions.  To obtain these, we take an even mode
linearised axisymmetric solutions for a given $l=l_w$, and rotate
this by $\pi/2$ so that the symmetry axis now lies in the
equator.  This generates a non-axisymmetric solution which is a
linear combination of all $l$ up to $l_w$ and all even $m$.
(This solution also satisfies the symmetry and regularity
conditions we have assumed).  Under this rotation, both even
and odd waves are generated.  The solution is
determined by an arbitrary wave amplitude $F(t-r)$ or $F(t+r)$
for the retarded or advanced solutions respectively.  Although
each solution is by itself singular at the origin, the combination
of (advanced $-$ retarded) solution is non-singular everywhere.
We choose a particular function $F$; construct initial data at
$t=0$; and evolve this initial data and observe the outgoing
waveform leaving the outer grid radius.  We then compare this
with the analytic solution.

\vskip .5 truein
\centerline {\bf II. \ Radial gauge and mixed slicing}

   We will not list here the full non-axisymmetric `3+1' equations
for our gauge choice.  These can be found in ref.[2] and
will be more fully discussed in a forthcoming paper [1].
We give here only the briefest details in order to 
define the variables used in our wave solution.

   We use spherical spatial coordinates ($r$,$\theta$,$\phi$). The 
spatial metric $h_{ij}$ ($i,j=1,2,3$) in the radial gauge [1-4] 
is determined from the conditions that
the off-diagonal ($r$,$\theta$) ($r$,$\phi$) components of the metric
be zero, and
that the angular spatial determinant be fixed to its Euclidean value.  
Thus we have:
     $$       h_{r\theta}=0     ,\eqno(1a)$$
     $$       h_{r\phi}=0       ,\eqno(1b)$$
$$ (h_{\theta\theta}h_{\phi\phi}-h^2_{\theta\phi})=r^4\sin^2\theta  .\eqno(1c)$$
The 3-dimensional metric is then determined by the three functions $A$, 
$\eta$, and $\xi$, each functions of ($t$, $r$, $\theta$, $\phi$), and the 
full 4-dimensional metric has the form:
   $$ds^2=-(N^2-N^iN_i)dt^2-2N_idx^idt+
          A^2dr^2+r^2B^{-2}d\theta^2
          +r^2 B^2(\sin\theta d\phi+\xi d\theta)^2 ,\eqno(2)$$                                                                 
\noindent with $B^2=1+\eta$, and where $\eta$ and $\xi$ are respectively
the even and odd dynamical degrees of freedom in the sense that they
tend (for this linearised test, and within a sign) at large radii to 
the even and odd transverse 
traceless amplitudes $h_+$,$h_\times$.  $N$ is the lapse; $N^i$ the 
shift vector raised and lowered using $h_{ij}$.
    
  The components of the extrinsic curvature we use are based
on the orthonormal components $K_{(i)(j)}$ ($i,j=1,2,3$) 
which are the components
of $K_{ij}$ projected onto the orthonormal triad of basis vectors:
$$e^i_{(1)}=[A^{-1},0,0]; \quad  e^i_{(2)}=[0,B/r,-\xi B/(r\sin\theta)]; \quad  
e^i_{(3)}=[0,0,1/(Br\sin\theta)]   .\eqno(3)$$ 

  The computed geometrical variables are (we do not discuss here
the accompanying hydrodynamic variables):
$$ {\rm Spatial \ metric:}  \qquad A;  \qquad  \eta; \qquad \xi   ;\eqno(4a)$$
 Extrinsic curvature: \qquad $K_1=K_{(1)(1)}$; 
                     \qquad  $K_2=BK_{(1)(2)}/\sin\theta$; 
   $$                  K_3=BK_{(1)(3)}/\sin\theta;
\qquad K_+={1\over 2}B^2(K_{(3)(3)}-K_{(2)(2)}); \qquad 
                                 K_\times=B^2K_{(2)(3)}  ;\eqno(4b)$$
$$ {\rm Lapse \ and \ shift:}  \qquad N; \qquad \beta^r=N^r/r; 
               \qquad G=N^\theta/\sin\theta;  \qquad N^\phi  .\eqno(4c)$$.

   Following ref.[4] we use a mixed foliation condition consisting of a
linear combination (depending on radius) of polar and maximal slicing: 
we set                                                                                 
        $${\rm trace}(K_i{}^j)=(1-C(r))K_r{}^r ,\eqno(5)$$                                                        
where $0\leq C(r)\leq 1$ is a chosen smooth function 
of radius such that $C(0)=1$ and $C(r>r_0)=0$ (corresponding to maximal 
slicing at the origin and polar slicing for                   
radii larger than a chosen transition radius $r_0$).
The actual form of $C(r)$ used is:
                   $$C(r)=[1-(r/r_0)^2]^n  ,\eqno(6)$$ 
with \ $r_0=r_0(t)$, $n$ chosen values.

   Regularity at the origin is determined by requiring  
the components in the `cartesian' ($\x$,$\y$,$\z$)
system of coordinates be expandable in non-negative integer 
powers of ($\x$,$\y$,$\z$) (consistent with the symmetry assumed) [4].
($\x$,$\y$,$\z$ are quasi-cartesian
coordinates related to $r$,$\theta$,$\phi$ in the usual manner).
This gives, in the linearised case (with no mean origin azimuthal
shift rotation), the 
following values for the origin shift and extrinsic curvature: 
$$\beta^r(r=0)=\lbrace k_+(3\cos^2\theta-1)
                    +(k_-\cos{2\phi}-k_\times\sin{2\phi})
                                \sin^2\theta\rbrace  ,\eqno(7a)$$
$$G(r=0)=\lbrace -3k_+-k_\times\sin{2\phi}+k_-\cos{2\phi}
                         \rbrace \cos\theta  ,\eqno(7b)$$
$$N^\phi(r=0)=\lbrace -k_-\sin{2\phi}-k_\times\cos{2\phi}
                         \rbrace  ,\eqno(7c)$$
$$K_+(r=0)={3\over 2}k_+\sin^2\theta+{1\over 2} (1+\cos^2\theta)[-k_\times\sin{2\phi}+k_-\cos{2\phi}] ,\eqno(8a)$$
$$K_\times(r=0)=\cos\theta[k_-\sin{2\phi}+k_\times\cos{2\phi}] ,\eqno(8b)$$
$$K_1(r=0)=k_+(1-3\cos^2\theta)+\sin^2\theta[-k_-\cos{2\phi}+k_\times\sin{2\phi}] ,\eqno(8c)$$
$$K_2(r=0)=\cos\theta[3k_+-k_-\cos{2\phi}+k_\times\sin{2\phi}] ,\eqno(8d)$$
$$K_3(r=0)=[k_-\sin{2\phi}+k_\times\cos{2\phi}] .\eqno(8e)$$
where $k_+(t)$, $k_-(t)$, $k_\times(t)$ are time dependent coefficients.

  Also we have near the origin:
 $$ \eta = O(r^2)  ,\eqno(9) $$
 $$ N = N(r=0)+O(r^2)  .\eqno(10) $$
     
\vskip .2 truein
\centerline {\bf III. \ Axisymmetric linearised gravitational waves}

   The axisymmetric solution we rotate is based on one found
by Bardeen [5,6] (see also ref.[7] where axisymmetric 
comparison tests are also shown).  Bardeen's solution corresponds 
to a general
$l$ even mode Teukolsky wave [8] transformed from the
transverse-traceless gauge to the radial gauge with maximal
slicing.  For completeness, we have generalised Bardeen's solution
(which is for maximal slicing) to the case of mixed slicing by obtaining
the linearised axisymmetric
equations in the radial gauge with mixed slicing.  As we explain below 
however, the original Bardeen solution is actually sufficient, even
for mixed slicing, for the comparison with the computed outgoing waves.

   Following Bardeen [5,6], we write the even linearised solution
for each $l\ge 2$ in the form:
      $$ (A-1)=\alpha_l P_{l0}  ,\eqno(11a)$$
 $$ \eta = [c_l + {{(a_l-2f_l)}\over{l(l+1)}}](1-x^2)P_{l0},_{xx}  ,\eqno(11b)$$
    $$ \xi =0  ,\eqno(11c)$$
    $$ K_1 = -{\dot a_l}P_{l0}  ,\eqno(11d)$$
    $$ K_2 = -{1\over 2}{\dot b_l}P_{l0},_x  ,\eqno(11e)$$
    $$ K_3 = 0  ,\eqno(11f)$$
    $$ K_+ = -{1\over 2}{\dot c_l}(1-x^2)P_{l0},_{xx}  ,\eqno(11g)$$
    $$ K_\times = 0  ,\eqno(11h)$$
    $$ G = {{(2{\dot f_l}-{\dot a_l})}\over{l(l+1)}}P_{l0},_x  ,\eqno(11i)$$
    $$ N^\phi = 0  ,\eqno(11j)$$
    $$ \beta^r = [-{\dot f_l}+{1\over 2}{\dot a_l}(1-C(r))]P_{l0} ,\eqno(11k)$$
    $$ N = n_lP_{l0}  ,\eqno(11l)$$
where \ $\cdot\equiv ({{\partial}\over{\partial t}})_r$.  In this form,
the evolution equation for $\eta$ [1,2] is automatically satisfied. 
The equation for $\beta^r$ [1,2] is also satisfied.

   The momentum constraints for $K_1$ and $K_2$ [1,2] require:
   $$ b_l=-{2\over{l(l+1)}}\lbrace r^{-2}[r^3a_l],_r
              -[r(1-C(r))a_l],_r \rbrace  ,\eqno(12)$$
   $$ c_l={1\over{(l-1)(l+2)}}\lbrace r^{-2}[r^3b_l],_r
               +a_l+(1-C(r))a_l \rbrace  .\eqno(13)$$
The shift equation for $G$ [1,2] provides an ordinary differential
equation for $f_l$:
   $$ 2rf_l,_r+l(l+1)f_l=-ra_l,_r-6a_l+2[r(1-C(r))a_l],_r
                     +{1\over 2}l(l+1)(1-C(r))a_l  .\eqno(14)$$
The origin boundary conditions (7),(8) (in the
axisymmetric limit) and (9), give for the origin boundary condition for
$f_l$:
   $$ f_l(r=0) = \cases{-a_l(r=0),&for $l=2$;\cr  
                            0,&for $l>2$\cr}  .\eqno(15)$$
The evolution equation for $A$ [1,2] gives:
 $$ \alpha_l =a_l+f_l+rf_l,_r-{1\over 2}[(1-C(r))ra_l],_r  ,\eqno(16)$$
or equivalently using (14) for $f_l$:
 $$ \alpha_l =-2a_l-{1\over 2}ra_l,_r-{1\over 2}(l-1)(l+2)f_l
   +{1\over 2}[(1-C(r))ra_l],_r+{1\over 4}l(l+1)(1-C(r))a_l  .\eqno(17)$$
The evolution equation for $K_1$ [1,2] gives (using (16),(17)):
 $$ r^2{\ddot a_l}=r^2n_l,_{rr}+r^2a_l,_{rr}+6ra_l,_r-(l+3)(l-2)a_l
      -[r^2(1-C(r))a_l],_{rr}  ,\eqno(18)$$
while the lapse equation [1,2], using (18), gives:
  $$ r^2n_l,_{rr}+2rn_l,_r-l(l+1)n_l=(1-C(r))r^2{\ddot a_l}  ,\eqno(19)$$
with origin boundary conditions (10):
  $$ n_l(r=0)=0; \ n_l,_r(r=0)=0  .\eqno(20)$$
The remaining equations: the Hamiltonian constraint [1,2], and the $K_2$
evolution equation [1,2], are then also satisfied by these solutions.

   For the case of maximal slicing ($C(r)=1$) we can explicitly solve
(18),(19) for $a_l$, $n_l$ and recover Bardeen's solution.  In this case,
the solution for $n_l$ is simply:
    $$ n_l=0  .\eqno(21)$$
The solution for $a_l$ for the case of a retarded solution (where $t$
occurs only in the form of retarded time $(t-r)$) is found using the
form:
    $$ a_l=\sum_n c_n r^{-(n+3)} F_l^{(l-n)}(t-r)  ,\eqno(22)$$
where $F_l^{(k)}$ indicates the $k$'th derivative with respect to the
arbitrary generating function $F_l$ for the $l$'th mode.  Substitution
into (18) gives for the coefficients $c_n$:
    $$ c_n=c_{n-1}{{(l+n)(l-n+1)}\over{2n}}  ,\eqno(23)$$
leading to the retarded solution:
 $$ a_l=\sum_{n=0}^l {{(l+n)!}\over{2^nn!(l-n)!}}
                r^{-(n+3)}F_l^{(l-n)}(t-r)  .\eqno(24)$$
The advanced solution is:
 $$ a_l=\sum_{n=0}^l {{(l+n)!}\over{2^nn!(l-n)!}}(-1)^n
                r^{-(n+3)}F_l^{(l-n)}(t+r)  ,\eqno(25)$$
while the non-singular (advanced $-$ retarded) solution is:
 $$ a_l=\sum_{n=0}^l {{(l+n)!}\over{2^nn!(l-n)!}}
     r^{-(n+3)}[(-1)^nF_l^{(l-n)}(t+r)-F_l^{(l-n)}(t-r)]  .\eqno(26)$$

  Substituting the non-singular solution (26) for $a_l$ into (14) 
(with $C(r)=1$) we find for $f_l$:
 $$ f_l=\sum_{n=0}^l {{(l+n)!}\over{2^{n+1}n!(l-n)!}}
     r^{-{1\over 2}l(l+1)}\int_0^r dr r^{[{1\over 2}l(l+1)-n-4]}\lbrace $$
  $$    -r[(-1)^nF_l^{(l-n+1)}(t+r)+F_l^{(l-n+1)}(t-r)] 
    +(n-3)[(-1)^nF_l^{(l-n)}(t+r)-F_l^{(l-n)}(t-r)] \rbrace  .\eqno(27)$$
The integral can be evaluated either directly, or by reducing
it using integration by parts to recurrence relations involving
the $F_l^{(k)}(t\pm r)$.

   The outgoing waveform at asymptotically large radii of the transverse
traceless amplitudes is:
  $$ h_+ = -{2\over{(l-1)l(l+1)(l+2)}}r^{-1}F^{(l+2)}(t-r) 
        (1-x^2)P_{l0},_{xx}  \quad (r\rightarrow \infty)     ,\eqno(28) $$
  $$ h_\times = 0   .\eqno(29)$$

\vskip .5 truein
\centerline {\bf IV. \ Generation of non-axisymmetric linearised 
gravitational waves}

   We now rotate this solution,
with the symmetry axis of the wave rotated into the 
equator, to generate a non-axisymmetric solution. 
The relation between
($\theta$,$\phi$) to the rotated ($\theta^\prime$,$\phi^\prime$) coordinates 
aligned with the symmetry axis of the wave (which when
rotated lies in the equator at an angle $\alpha$ to the $\x$ axis
of the ($\theta$,$\phi$) system, 
corresponding to $\x^\prime =-\z$; $\y^\prime =\y\cos\alpha-\x\sin\alpha$; 
$\z^\prime =\x\cos\alpha+\y\sin\alpha$; $r^\prime =r$) is:
 $$ \sin\theta^\prime \cos\phi^\prime =-\cos\theta  ,\eqno(30a)$$
 $$ \sin\theta^\prime \sin\phi^\prime = \sin\theta \sin (\phi-\alpha)  
 ,\eqno(30b)$$
 $$ \cos\theta^\prime = \sin\theta \cos (\phi-\alpha)  .\eqno(30c)$$
Under this rotation, to linear order, the geometrical variables become:
 $$ \eta (t,r,\theta,\phi)={{\eta^\prime (t,r,\theta^\prime)}\over
   {\sin^2\theta^\prime}}[\cos{2(\phi-\alpha)}
        -{1\over 2}\sin^2\theta (1+\cos{2(\phi-\alpha)})]  ,\eqno(31a) $$
 $$ \xi (t,r,\theta,\phi)={{\eta^\prime (t,r,\theta^\prime)}\over
   {\sin^2\theta^\prime}} \cos\theta \sin{2(\phi-\alpha)}  ,\eqno(31b)$$
 $$ G(t,r,\theta,\phi)=-{{G^\prime (t,r,\theta^\prime)}\over{\cos\theta^\prime}}
        \cos\theta {1\over 2}(1+\cos{2(\phi-\alpha)})  ,\eqno(31c)$$
 $$ N^\phi (t,r,\theta,\phi)={{G^\prime (t,r,\theta^\prime)}
         \over{\cos\theta^\prime}}{1\over 2}\sin{2(\phi-\alpha)}  ,\eqno(31d)$$
 $$ \beta^r(t,r,\theta,\phi)={\beta^{r\prime} (t,r,\theta^\prime)} ,\eqno(31e)$$
 $$ A(t,r,\theta,\phi)={A^\prime (t,r,\theta^\prime)}  ,\eqno(31f)$$
 $$ N(t,r,\theta,\phi)={N^\prime (t,r,\theta^\prime)}  ,\eqno(31g)$$
 $$ K_+(t,r,\theta,\phi)={{K_+^\prime (t,r,\theta^\prime)}
          \over{\sin^2\theta^\prime}}{1\over 2}
        [-\sin^2\theta +(1+\cos^2\theta)\cos{2(\phi-\alpha)}]  ,\eqno(31h)$$
 $$ K_\times (t,r,\theta,\phi)={{K_+ ^\prime (t,r,\theta^\prime)}
      \over{\sin^2\theta^\prime}}\cos\theta\sin{2(\phi-\alpha)}  ,\eqno(31i)$$
 $$ K_1(t,r,\theta,\phi)={K_1^\prime (t,r,\theta^\prime)}  ,\eqno(31j)$$
 $$ K_2(t,r,\theta,\phi)={{-K_2^\prime (t,r,\theta^\prime)}
          \over{\cos\theta^\prime}}{1\over 2}\cos\theta
              (1+\cos{2(\phi-\alpha)})  ,\eqno(31k)$$
 $$ K_3(t,r,\theta,\phi)={{K_2^\prime (t,r,\theta^\prime)}
      \over{\cos\theta^\prime}}{1\over 2}\sin{2(\phi-\alpha)}  .\eqno(31l)$$

    Applying these rotated values to the non-singular solution (26),(27)
and expanding around the origin, we find the following values for the 
origin coefficients
$k_+(t)$, $k_-(t)$, $k_\times(t)$ in  (7),(8) when $l=2$: 
     $$ k_+ = -{1\over 30} F^{(6)}_{l=2}(t)|_{r=0}  ,\eqno(32a)$$
     $$ k_- = {1\over 10} F^{(6)}_{l=2}(t)|_{r=0}\cos 2\alpha  ,\eqno(32b)$$
   $$ k_\times = -{1\over 10} F^{(6)}_{l=2}(t)|_{r=0}\sin 2\alpha  .\eqno(32c)$$
For $l>2$ these coefficients are zero.

   The $l=2$ solution after rotation
corresponds to the following linear combination of waves:
  $$ \eta ={\tilde \eta}_2(r,t)[-{1\over 2}P_{22}
               +(3-{1\over 2}P_{22})\cos 2(\phi-\alpha) ]  ,\eqno(33a)$$
  $$ \xi = {\tilde \eta}_2(r,t) 3 P_{10}\sin 2(\phi-\alpha)  ,\eqno(33b)$$
  The $l=4$ rotated solution corresponds to the linear combination:
  $$ \eta ={\tilde \eta}_4(r,t)\lbrace {3\over 8}P_{42}
  +(-{15\over 2}+{5\over 2}P_{22}+{1\over 2}P_{42})\cos 2(\phi-\alpha)
  +({5\over 2}P_{22}+{1\over 8}P_{42})\cos 4 (\phi-\alpha)\rbrace  ,\eqno(34a)$$
  $$ \xi = {\tilde \eta}_4(r,t)
        \lbrace (-{{15}\over 2}P_{10}+{7\over 4}P_{32})\sin 2(\phi-\alpha)
           +{7\over 8}P_{32}\sin 4(\phi-\alpha) \rbrace  ,\eqno(34b)$$
where in the above: 
   $${\tilde \eta}_l(r,t)=[c_l+{{(a_l-2f_l)}\over{l(l+1)}}]  .\eqno(35)$$.

   Expanding $F_l$ near the origin, we find the leading order non-zero 
terms of ${\tilde \eta}_l$ for $l=2,4$ to be:
    $$ {\tilde \eta}_2=-{1\over {75}}r^2 F_{l=2}^{(7)}(t)|_{r=0} 
                               \quad (r\rightarrow 0)   ,\eqno(36) $$
    $$ {\tilde \eta}_4=-{{1}\over {79380}}r^4 F_{l=4}^{(9)}(t)|_{r=0} 
                               \quad (r\rightarrow 0)   .\eqno(37) $$
This leading order behaviour is consistent with our assumption of
cartesian integer expandability near the origin.

  The outgoing form for ${\tilde \eta}_l(r,t)$ at asymptotically large
radii is:
  $$ {\tilde \eta}_l(r,t)={2\over{(l-1)l(l+1)(l+2)}}r^{-1}F^{(l+2)}_l(t-r)
              \quad (r\rightarrow \infty)               .\eqno(38) $$
Substitution of (38) into (33),(34) gives the outgoing waveform at 
asymptotically large radii. The transverse traceless amplitudes 
(to linear order) are then simply:
 $$ h_+ = -\eta  ,\eqno(39) $$
 $$ h_\times = \xi  .\eqno(40) $$

\vskip .5 truein
\centerline {\bf V. \ The numerically evolved solution}

   For the $l=2$ wave test, we choose the wave amplitude function
$F_l$ to be:
  $$ F_{(l=2)}(y)=a y\exp (-y^2)  .\eqno(41)$$
For $l=4$ we choose:
  $$ F_{(l=4)}(y)=a y^3\exp (-y^2)  ,\eqno(42)$$
where $a$ is the (small) amplitude of the wave.
The extra factor of $y^2$ in (42) is to simulate more complex radial
waveform structure.

   The moments $F_l^{(k)}$ in the solutions (26),(27) for our choices
(39),(40) for $F_l$ can be automatically generated essentially
analytically by repeated application of the recurrence relation:
   $$ y^p \exp(-y^2) \rightarrow
          (py^{p-1}-2y^{p+1}) \exp(-y^2)  ,\eqno(43)$$
for the mapping on differentiation of the coefficients multiplying
the powers of $y$.  Hence if required, the whole linearised solution 
(for maximal slicing) can be obtained essentially `analytically' 
throughout the full spacetime for comparison with numerical code solutions.

   An important point to note is that although our `analytic' solution
is for maximal slicing, we can still use it for comparison with the
outgoing solution at large radii obtained numerically with mixed slicing.  
This is because firstly, the choice (41),(42) corresponds at $t=0$ to
time symmetric initial data with ($K_1$,$K_2$,$K_3$,$K_+$,$K_\times$) all zero.
Hence initially the mixed and maximal slicing are equivalent.
Secondly during the evolution at large $r$ where we monitor the
outgoing wave, $h_+$,$h_\times$ are gauge invariant while the
coordinates times agree (differing only by terms $\sim a^2/r$).

   $h_+$ and $h_\times$ can be spectrally represented (to linear order)
in the following manner [1]: 
$$   h_+ =(-E_{\rm axis}\cos(2\phi)+D_{\rm axis}\sin(2\phi)) $$
  $$  -\sum_{l=2,4,6...}(a_l+a_{l0}\cos(2\phi)+b_{l0}\sin(2\phi))P_{l2} $$
     $$  -\sum_{{l=2,4,6,...}\atop {m=2,4,...,l}} 
             (a_{lm}\cos((m+2)\phi)+b_{lm}\sin((m+2)\phi))P_{lm}  ,\eqno(44a)$$
$$ h_\times = (D_{\rm axis}\cos(2\phi)+E_{\rm axis}\sin(2\phi))P_1 $$
  $$  +\sum_{l=3,5,7...}(a^\prime_l+a^\prime_{l0}\cos(2\phi)
                        +b^\prime_{l0}\sin(2\phi))P_{l2} $$
    $$           +\sum_{{l=3,5,7...}\atop{m=2,4,..,l-1}}
             (a^\prime_{lm}\cos((m+2)\phi)
                      +b^\prime_{lm}\sin((m+2)\phi))P_{lm}  ,\eqno(44b)$$
where the spectral amplitudes 
$D_{\rm axis}$,$E_{\rm axis}$; $a_{l}$; $a_{l0}$,$b_{l0}$; $a_{lm}$,$b_{lm}$ 
and their primed counterparts,
are functions of radius and time.  These expansions are used to 
spectrally resolve the angular dependence of the numerically obtained
outgoing waves on the angular grid points at each radius.

   For the numerical evolution, we have used an outer grid radius of 
$10$ units; a grid of
(100x11x16) ($r,\theta ,\phi$) grid points; and an amplitude
$a=10^{-5}$ for $l=2$ and $a=10^{-8}$ for $l=4$. $r_0$ is 
$1.4534$ units, and $n=2.5$ (see eq.(6)).  We have taken 
$\alpha=0$.  (Details of the numerical code will be given in a
separate paper [9]). The expected non-zero asymptotic 
amplitudes (for these values) are, for $l=2$:
   $$ [E_{\rm axis},a_2,a_{20}]= 
     {1\over 24} \ [6,-1,-1] \ r^{-1}F^{(4)}_{(l=2)}(t-r) 
                          \quad (r\rightarrow \infty)  ,\eqno(45a)$$
where:  $$ F^{(4)}_{(l=2)}(y) = a \exp(-y^2)(60y-80y^3+16y^5)
                                  ,\eqno(45b)$$
while for $l=4$:
   $$ [E_{\rm axis},a_4,a_{20},a_{40},a_{22},a_{42},b'_{30},b'_{32}]= 
   {1\over {1440}} \ [-60,3,20,4,20,1,14,7] 
       \eqno r^{-1}F^{(6)}_{(l=4)}(t-r) \ \ (r\rightarrow \infty) \ \ (46a)$$
with:  $$ F^{(6)}_{(l=4)}(y) = a \exp(-y^2)
             (2520y-7560y^3+5040y^5-1056y^7+64y^9)  .\eqno(46b)$$

   Figures 1-11 show a comparison of the numerically computed non-zero 
spectral outgoing amplitudes versus the theoretically expected asymptotic 
waveform amplitudes  (45),(46).  Figures 1-3 show results for the 
$l=2$ test; figures 4-11 
results for the $l=4$ test.  The numerical waveforms are measured at the 
outer grid radius of $10$ units.  Reasonable agreement 
is obtained, although the later part of the numerical waveforms
tend to damp too slowly and eventually show irregularity.  The 
present preliminary setup of our code is unable to run much further 
than shown because of stability difficulties.  
Further work is currently underway to overcome these problems.   

   The tests presented here are derived from rotated axisymmetry
solutions, and are not the most general non-axisymmetric test possible.
More generally, radial gauge coordinates satisfying the origin
regularity conditions will not have all metric and shift variables
which explicitly fall-off asymptotically at large radii [1].
These more general types of tests will be dealt with in a separate
paper [9].

\vskip .5 truein
\centerline {\bf Acknowledgements}
\parindent=15pt

  A grant under the International Centre for Theoretical
Physics (ICTP)/Slovenia cooperation program sponsored by the Slovenian
Ministry of Science is gratefully acknowledged (R.F.S.).
One of us (R.F.S.) is especially grateful to Prof. Andrej Cadez
for his keen interest, encouragement and support.
This work has also been supported by an EEC fellowship (R.F.S.) at
the Scuola Internazionale Superiore di Studi Avanzati (SISSA)
within the Human Capital and Mobility research programme.
We are especially grateful to Prof. Dennis Sciama for the support
of this fellowship, as well as additional SISSA and ICTP fellowship
support.  One of us (R.F.S.) thanks Drs. John Miller and Antonio
Lanza for their interest.  We also thank Alvise Nobile, Head
of the Computer Center of the ICTP, and his staff for their assistance.
Use of the ICTP computational facilities, as well as those at the
Department of Physics, University of Ljubljana are acknowledged.

\baselineskip=11.25pt
\parskip=5pt plus 1pt
\vskip .5 truein
\centerline {\bf References:}
\vskip .2 truein
\parindent=0pt

1. L. Sigalotti and R.F. Stark, Non-axisymmetric rotating
gravitational collapse and gravitational radiation: theoretical
background, to be submitted to {\sl Phys.Rev.}D.. 

2. R.F. Stark, 1989, in {\sl Frontiers in numerical
relativity}, Eds. C.R.Evans, L.S.Finn\hfill\break 
\& D.W.Hobill, (Cambridge University Press).

3. D.M. Eardley, 1980, Unpublished working notes.

4. J.M. Bardeen and T. Piran, 1983, {\sl Physics Reports}, {\bf 96}, 205. 

5. J.M. Bardeen, 1983, unpublished private communication.

6. J.M. Bardeen, 1983, in {\sl Gravitational Radiation}, eds. N.Deruelle and
T.Piran, (North-Holland, Amsterdam).

7. R.F. Stark and T. Piran, 1987, {\sl Computer Physics Reports}, {\bf 5}, 221.

8. S.A. Teukolsky, 1982, {\sl Phys.Rev.}D., {\bf 26}, 745.

9. L. Sigalotti and R.F. Stark, Non-axisymmetric rotating
gravitational collapse and gravitational radiation: numerical
methods and tests, in preparation. 

\vskip .5 truein

\baselineskip=11.25pt
\parskip=9pt plus 1pt
\centerline {\bf Figure captions:}
\vskip .2 truein
\parindent=0pt

Fig.1: $l=2$ wave test; $E_{\rm axis}$ amplitude (see eqs.(44),(45)) 
vs. retarded time $(t-r)$; numerical and theoretical curves.
Comparison of the measured outgoing amplitude (full curve) 
from the full numerical code with the theoretical linearised 
prediction (broken curve).  
The amplitude is normalised by the small initial amplitude $a$.  
Show is the variation of the amplitude vs. retarded time.
The numerical solution is obtained at the outer grid radius of
10 units.  The theoretical curve is that for asymptotically
large radii.

Fig.2: $l=2$ wave test; $a_2$ amplitude (see eqs.(44),(45)) 
vs. retarded time $(t-r)$; numerical and theoretical curves.
(see also Fig.1 caption).

Fig.3: $l=2$ wave test; $a_{20}$ amplitude (see eqs.(44),(45)) 
vs. retarded time $(t-r)$; numerical and theoretical curves.
(see also Fig.1 caption).
  
Fig.4: $l=4$ wave test; $E_{\rm axis}$ amplitude (see eqs.(44),(46)) 
vs. retarded time $(t-r)$; numerical and theoretical curves.
(see also Fig.1 caption).
  
Fig.5: $l=4$ wave test; $a_4$ amplitude (see eqs.(44),(46)) 
vs. retarded time $(t-r)$; numerical and theoretical curves.
(see also Fig.1 caption).
  
Fig.6: $l=4$ wave test; $a_{20}$ amplitude (see eqs.(44),(46)) 
vs. retarded time $(t-r)$; numerical and theoretical curves.
(see also Fig.1 caption).
  
Fig.7: $l=4$ wave test; $a_{40}$ amplitude (see eqs.(44),(46)) 
vs. retarded time $(t-r)$; numerical and theoretical curves.
(see also Fig.1 caption).
  
Fig.8: $l=4$ wave test; $a_{22}$ amplitude (see eqs.(44),(46)) 
vs. retarded time $(t-r)$; numerical and theoretical curves.
(see also Fig.1 caption).
  
Fig.9: $l=4$ wave test; $a_{42}$ amplitude (see eqs.(44),(46)) 
vs. retarded time $(t-r)$; numerical and theoretical curves.
(see also Fig.1 caption).
  
Fig.10: $l=4$ wave test; $b'_{30}$ amplitude (see eqs.(44),(46)) 
vs. retarded time $(t-r)$; numerical and theoretical curves.
(see also Fig.1 caption).
  
Fig.11: $l=4$ wave test; $b'_{32}$ amplitude (see eqs.(44),(46)) 
vs. retarded time $(t-r)$; numerical and theoretical curves.
(see also Fig.1 caption).

\psfig {figure=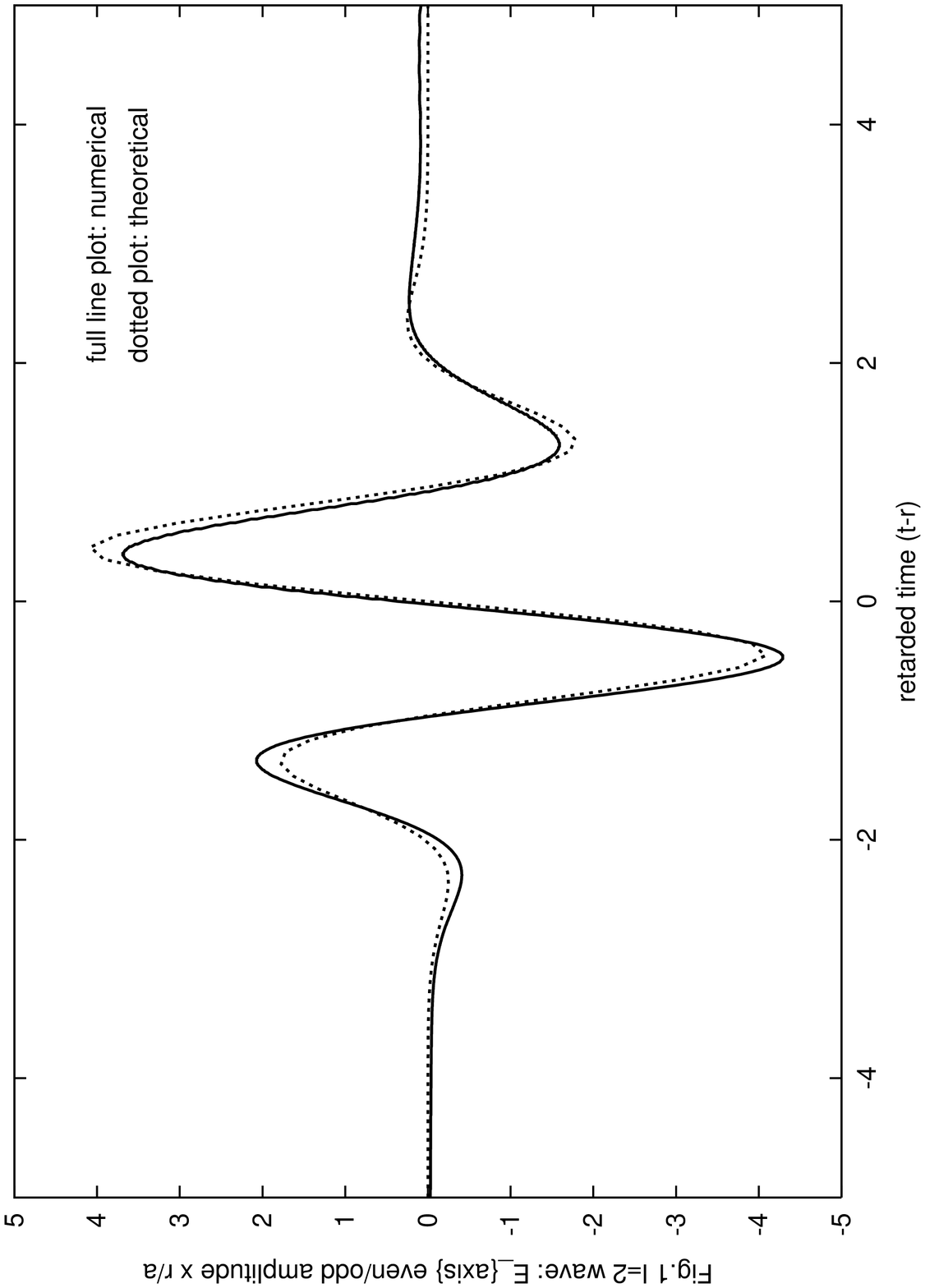,height=18cm}
\psfig {figure=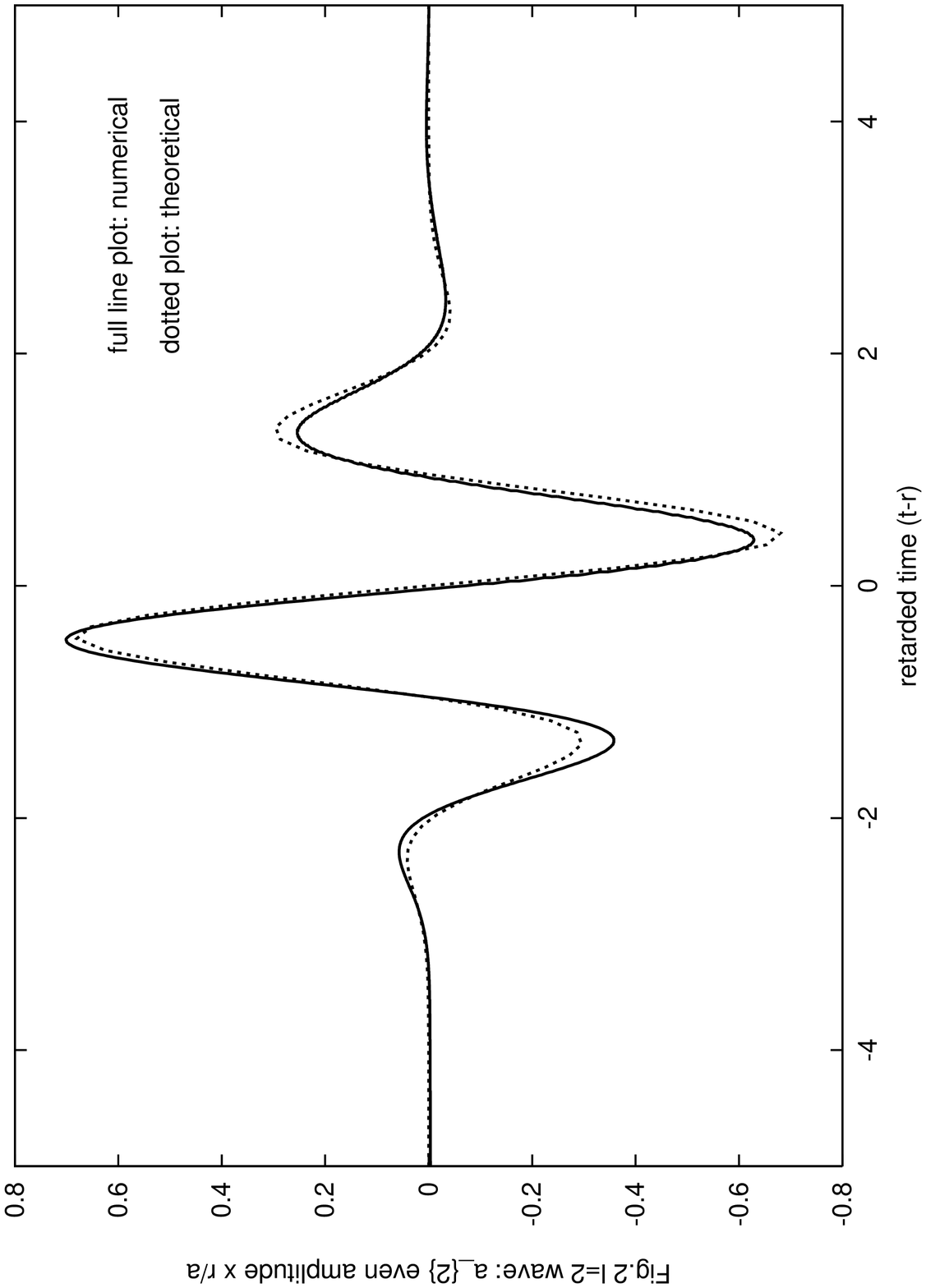,height=18cm}
\psfig {figure=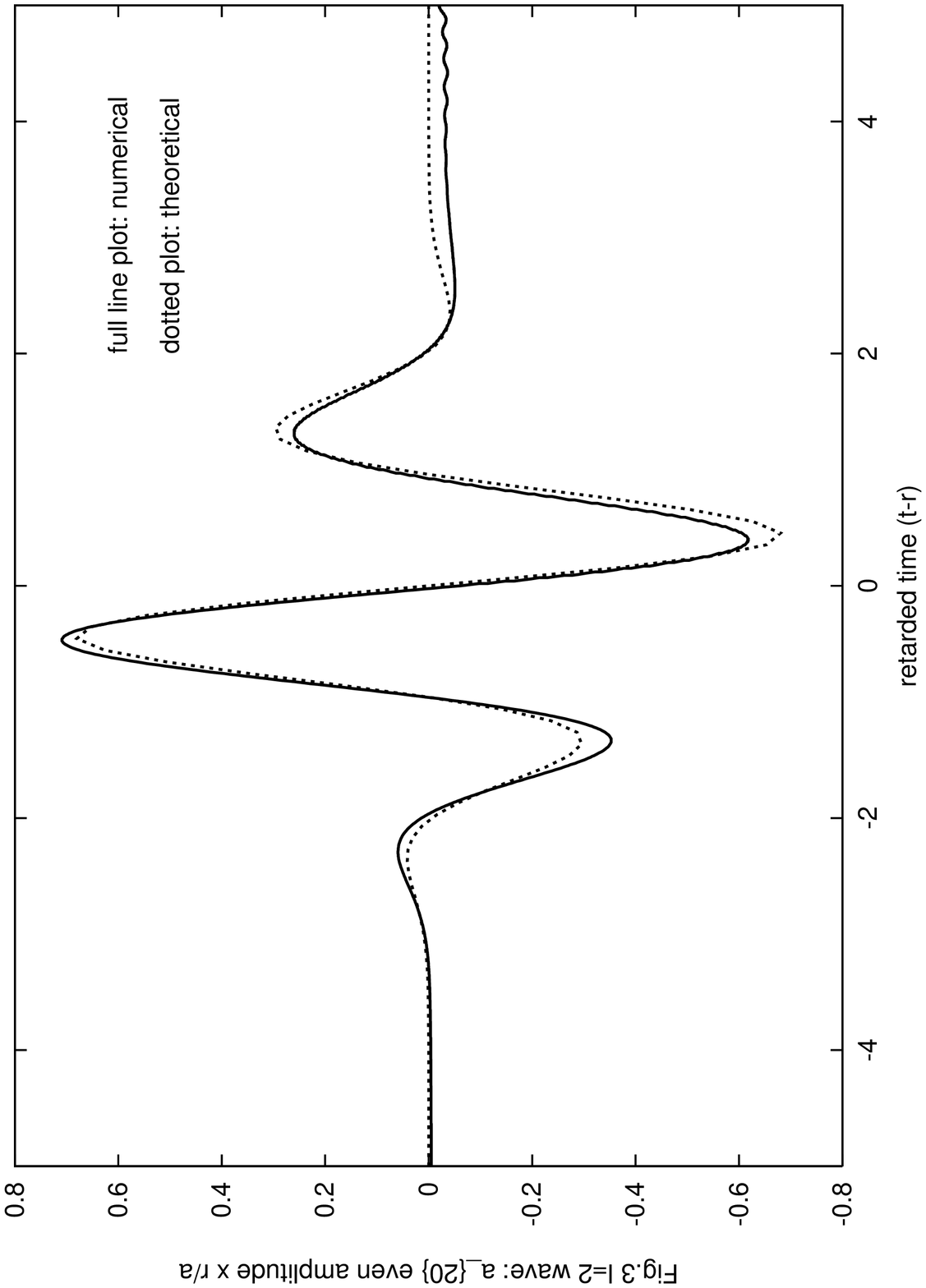,height=18cm}
\psfig {figure=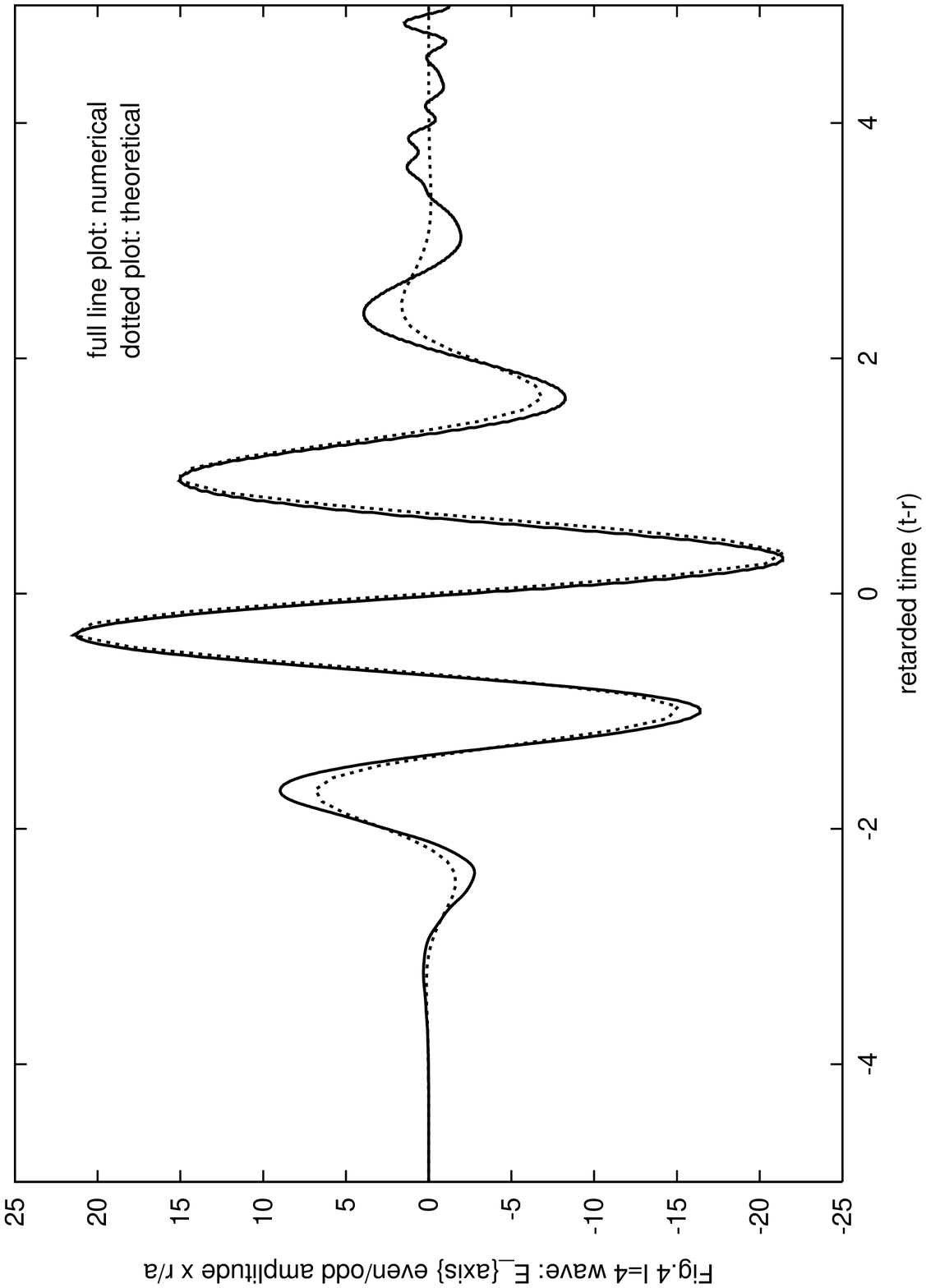,height=18cm}
\psfig {figure=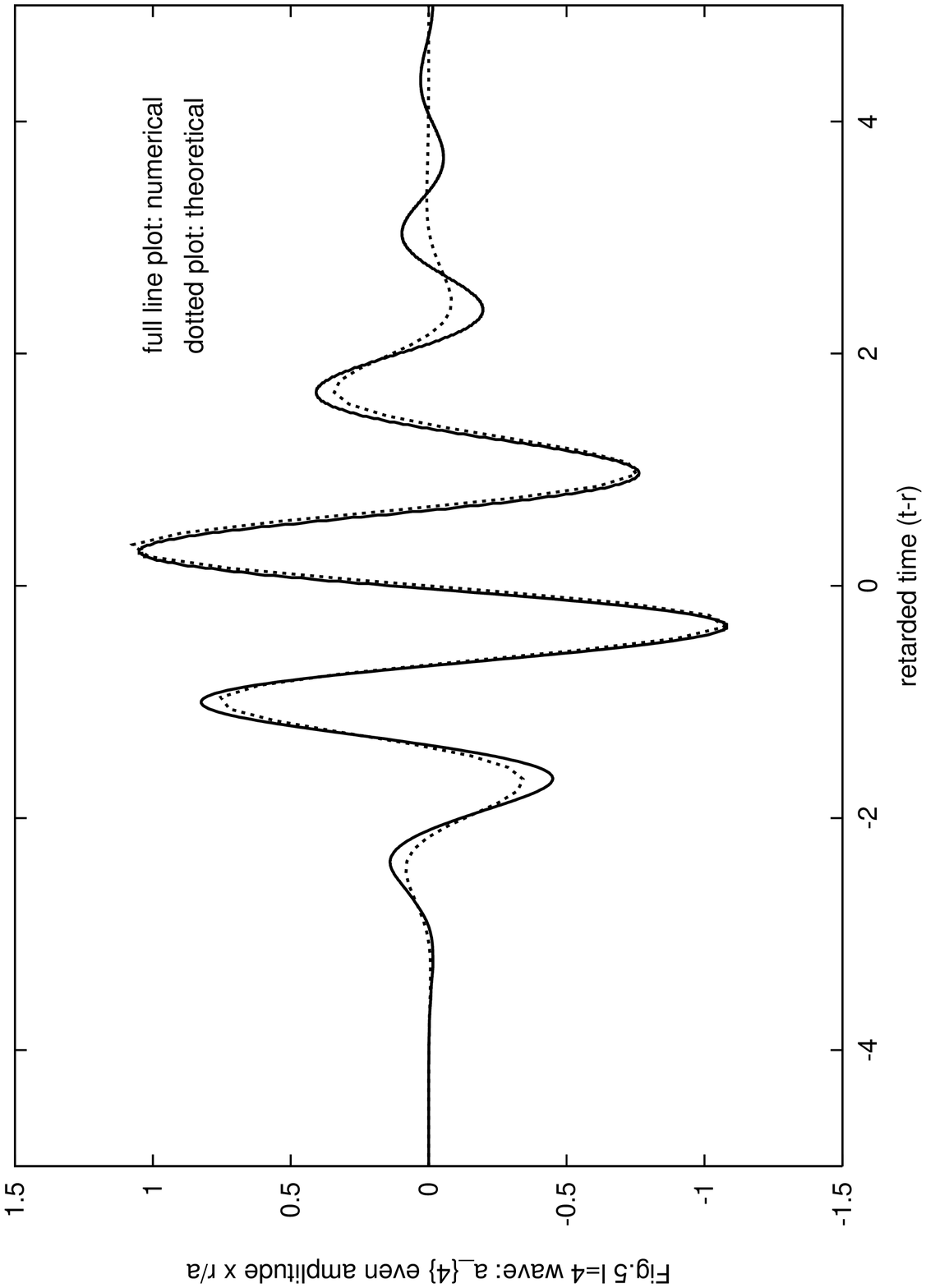,height=18cm}
\psfig {figure=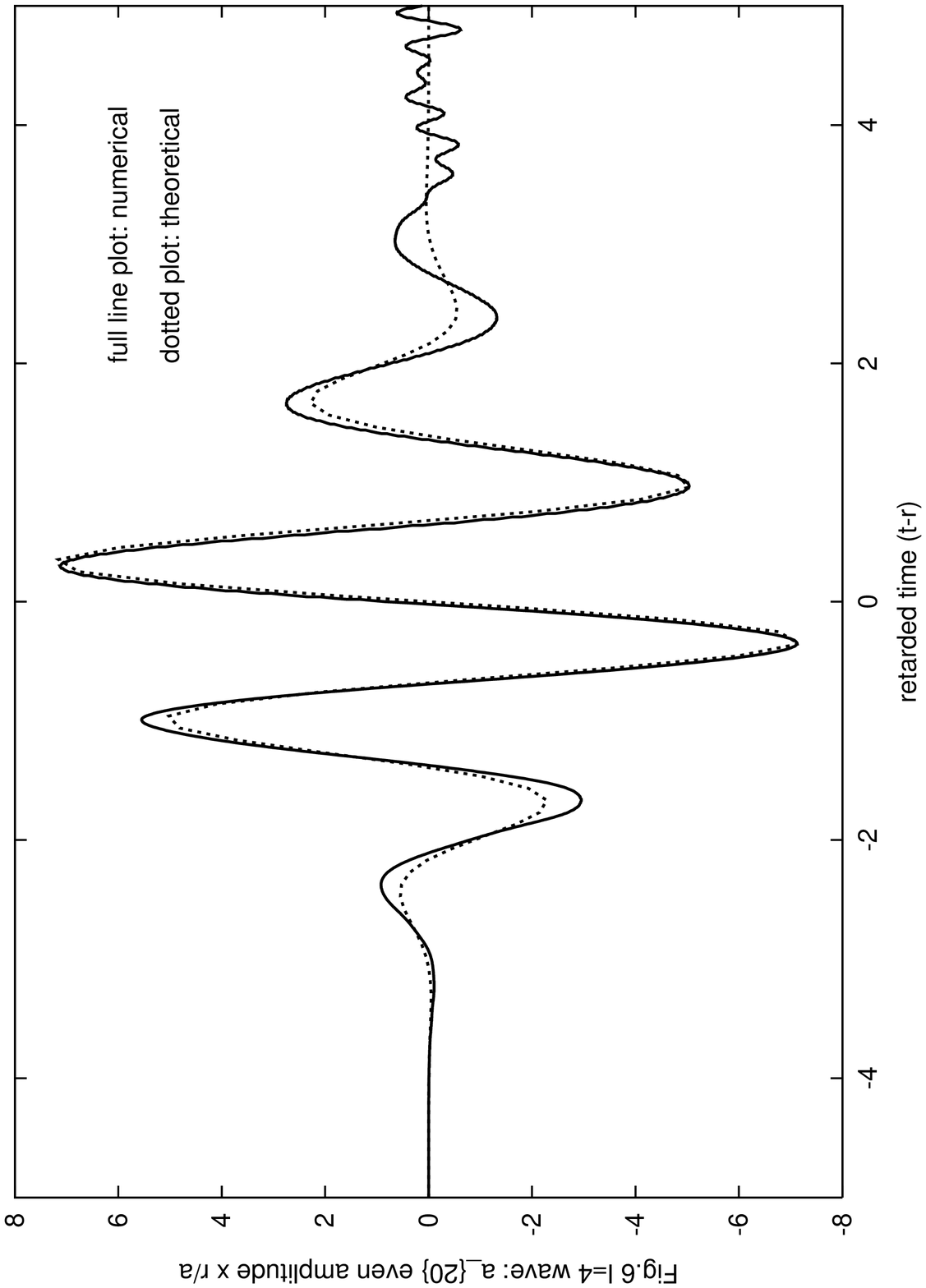,height=18cm}
\psfig {figure=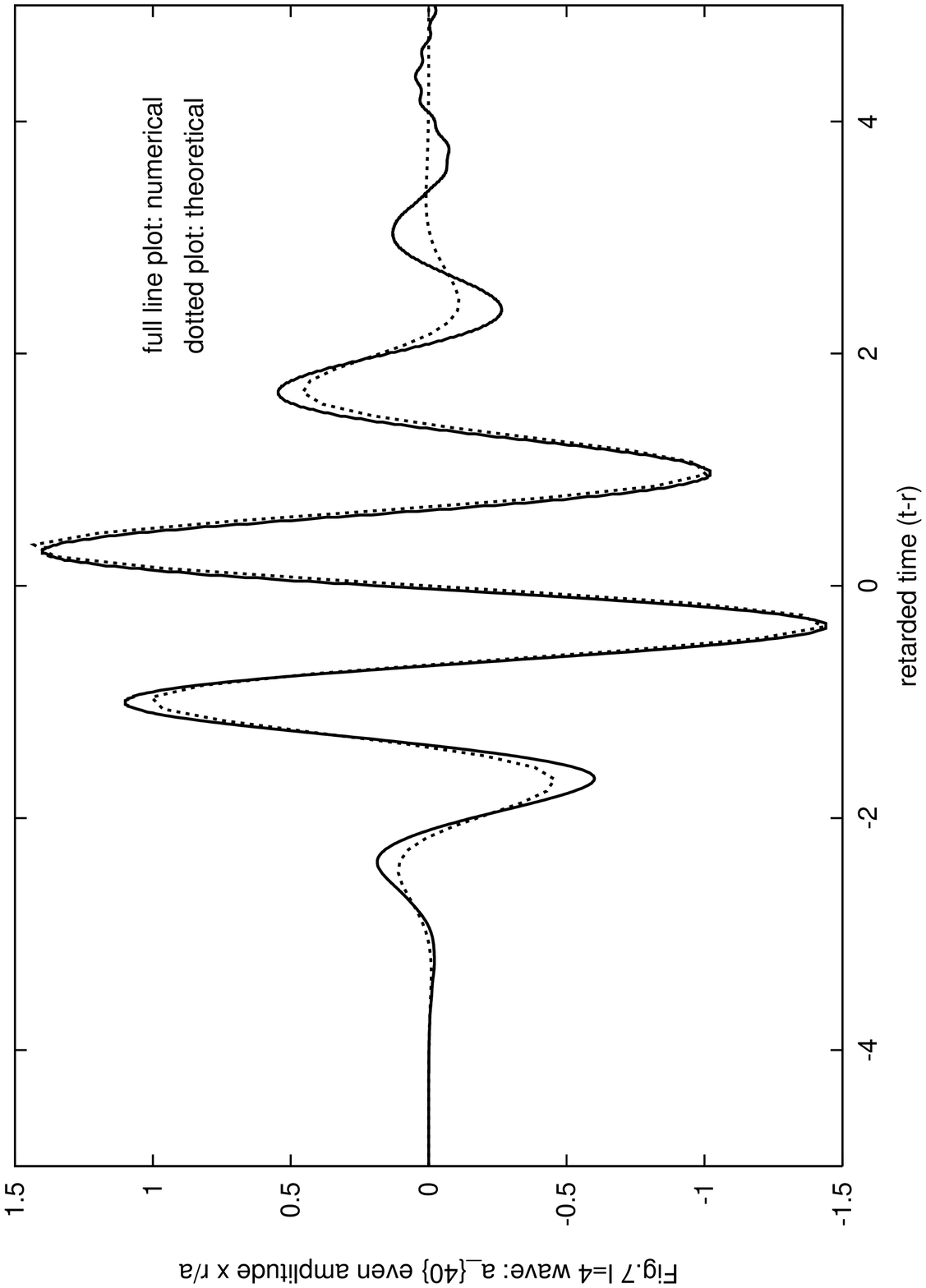,height=18cm}
\psfig {figure=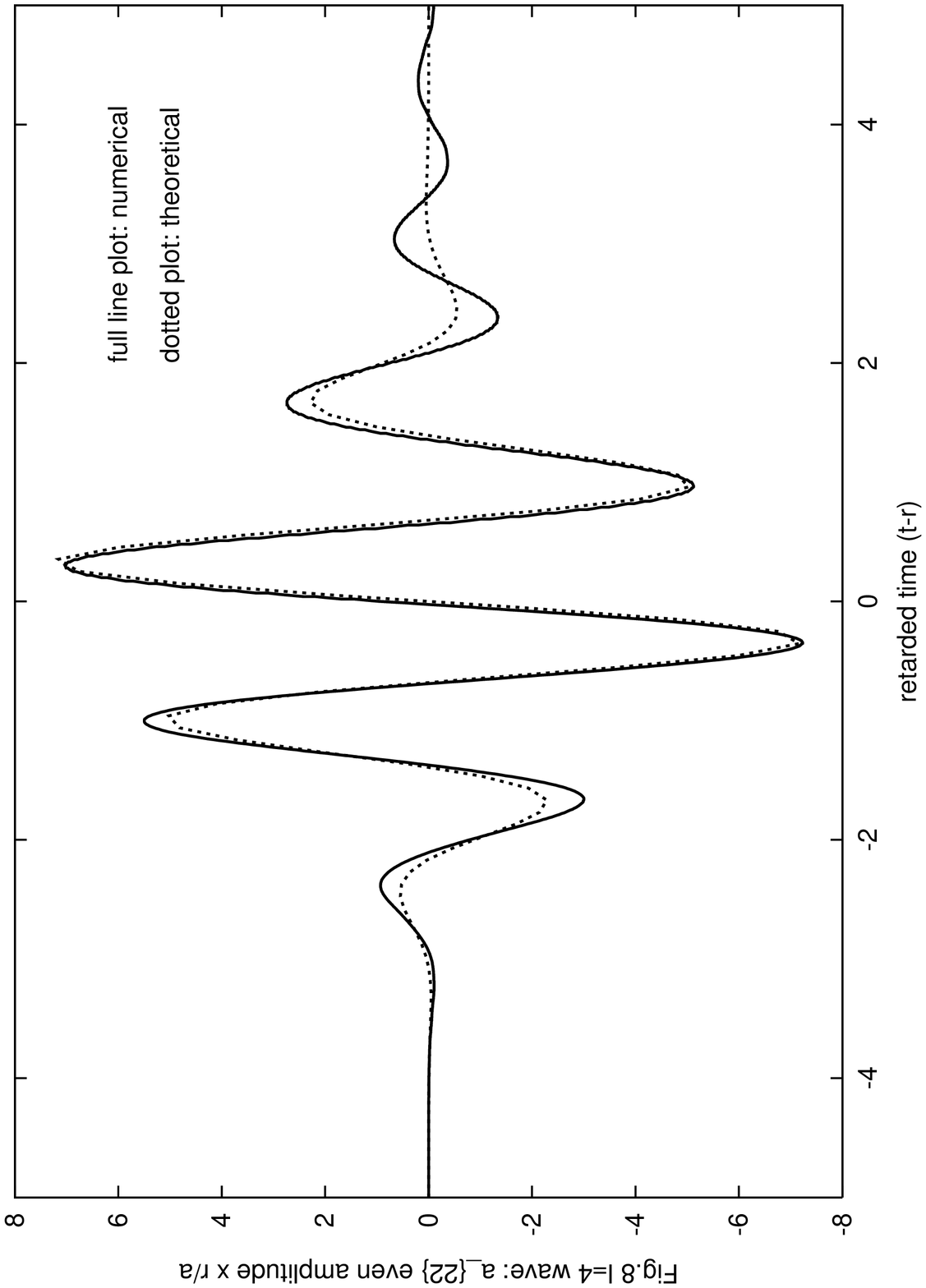,height=18cm}
\psfig {figure=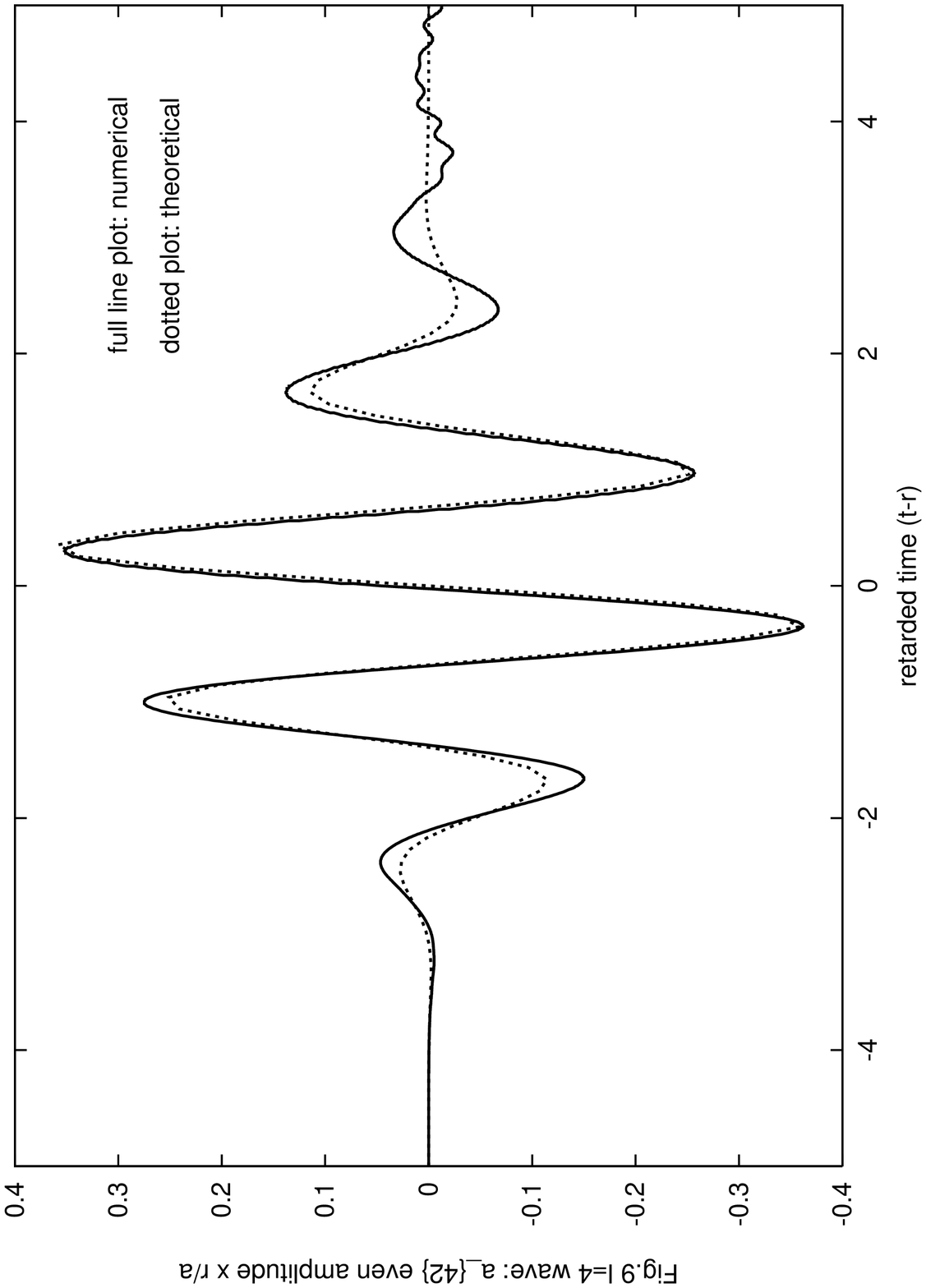,height=18cm}
\psfig {figure=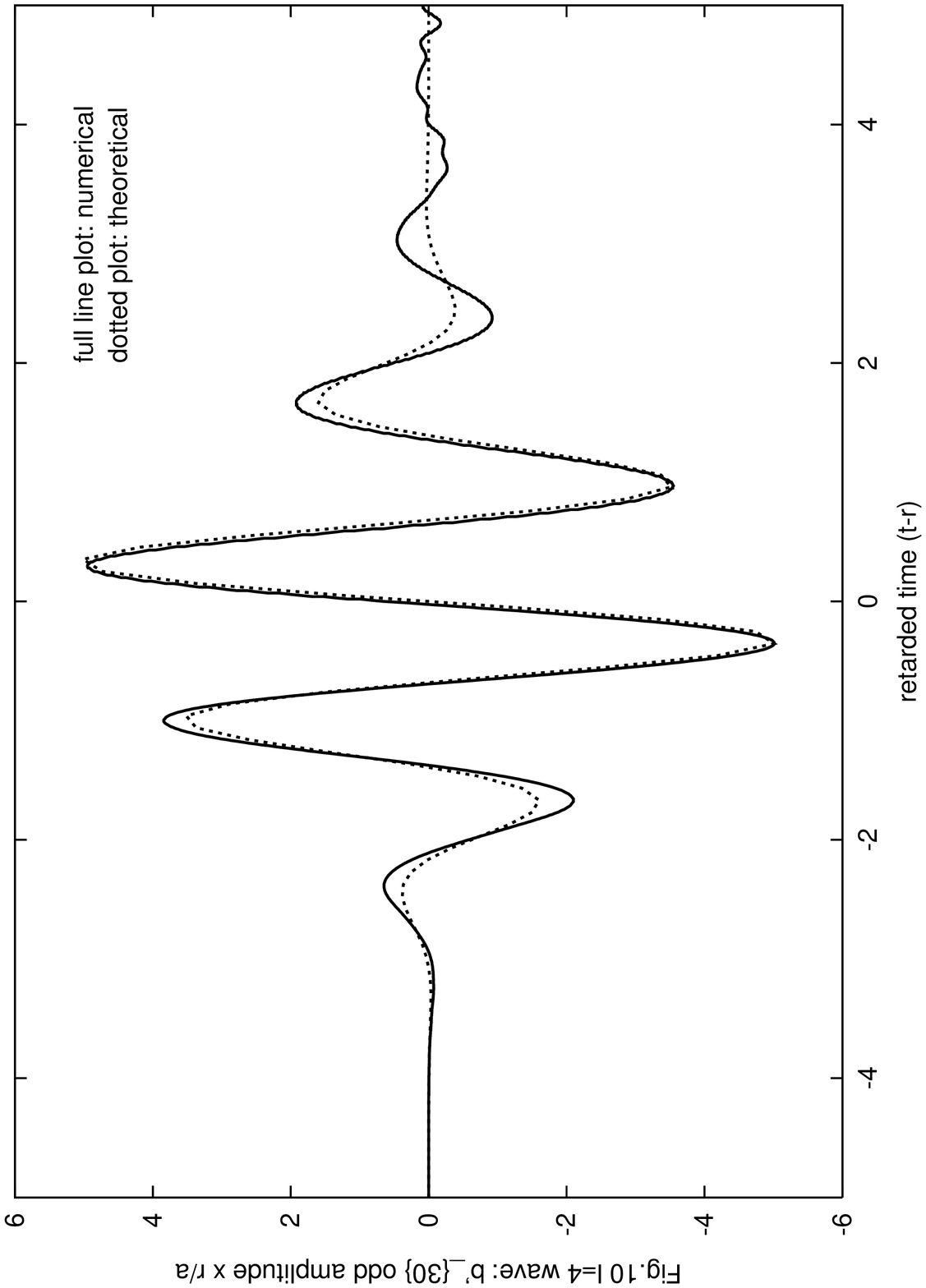,height=18cm}
\psfig {figure=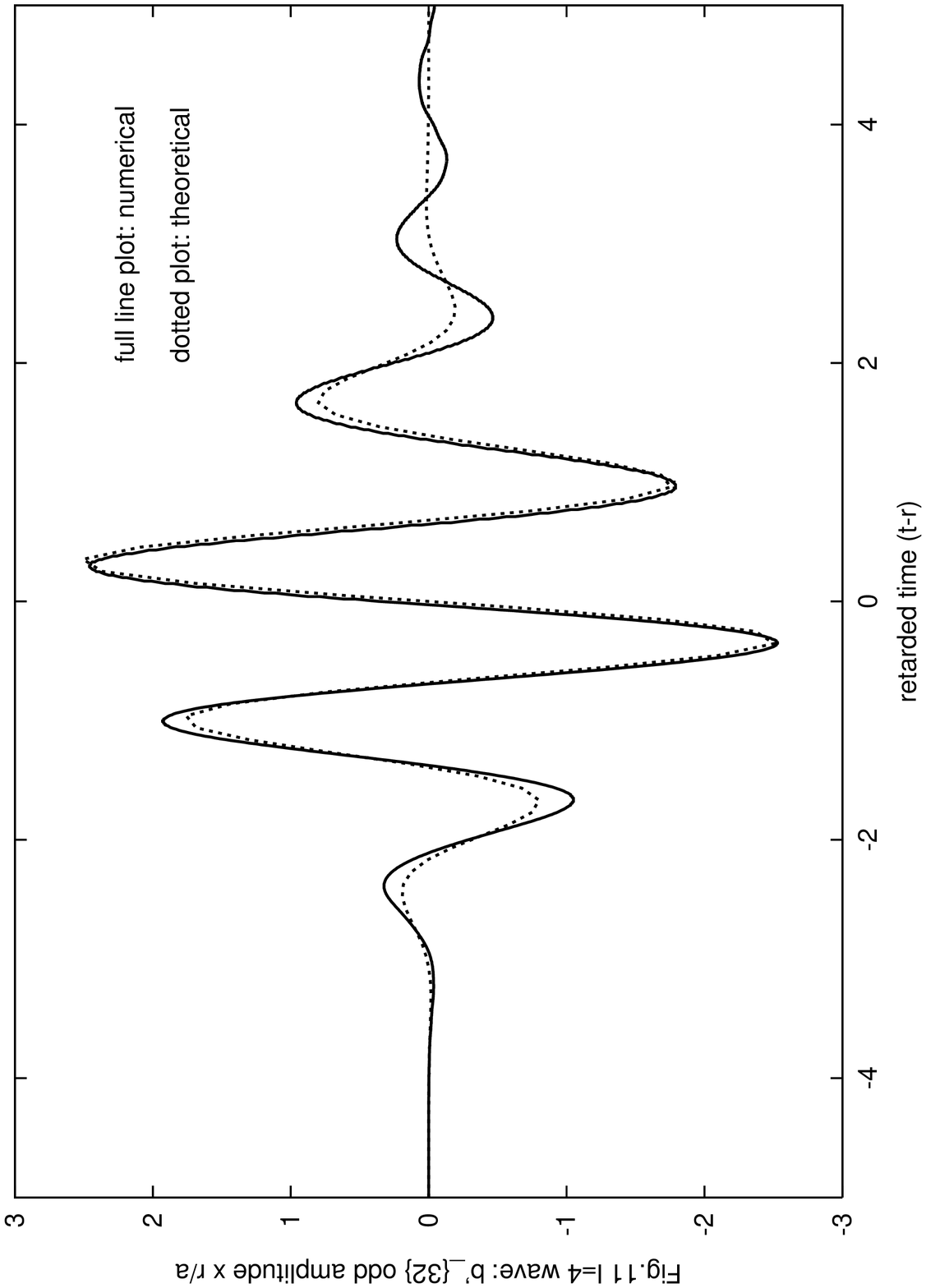,height=18cm}

\vfill
\end